\documentclass[
    ,final            
  ]
  {aipproc}

\layoutstyle{6x9}
\usepackage{graphicx}
\usepackage{psfrag}

\begin{document}

\title{Studying the infrared behaviour of gluon and ghost propagators using large asymmetric lattices}


\author{P. J. Silva}{
  address={Centro de F\'{i}sica Computacional, Departamento de F\'{i}sica, Universidade de Coimbra, P-3004-516 Coimbra, Portugal}
}

\author{O. Oliveira}{
  address={Centro de F\'{i}sica Computacional, Departamento de F\'{i}sica, Universidade de Coimbra, P-3004-516 Coimbra, Portugal}
}

\begin{abstract}
We report on the infrared limit of the
quenched lattice Landau gauge gluon propagator computed from large
asymmetric lattices. In particular, the compatibility of the pure power
law infrared solution $( q^2 )^{2\kappa}$ of the Dyson-Schwinger equations
is investigated and the exponent $\kappa$ is measured. Some results for
the ghost propagator and for the running coupling constant will also be
shown.
\end{abstract}

\maketitle


Despite the success of Quantum Chromodynamics (QCD) as \textit{the} theory of the 
strong interaction, a full understanding of the confinement mechanism is still 
missing. One line of research, very active in the last years, consists in the 
study of the QCD propagators for low momenta. Indeed, some works (for details, 
see \cite{AlkLvS01} and references therein) relate the infrared behaviour 
of the gluon 
and ghost propagators in the Landau gauge with gluon confinement. 
In particular, the Zwanziger horizon condition implies a null zero 
momentum gluon propagator $D(q^2)$, and the 
Kugo-Ojima confinement mechanism requires an infinite zero momentum ghost 
propagator $G(q^2)$.

An investigation of the infrared behaviour of the gluon 
and ghost propagators should be done in a non-perturbative framework. 
At the moment, two first principles approaches are available for such a task, 
namely Dyson-Schwinger equations (DSE) and lattice QCD methods. Given the 
different nature of such approaches, a comparison between the results of 
the two methods is necessary.

A solution of the DSE \cite{lerche} predicting pure power laws for gluon 
and ghost dressing functions, 
\begin{equation}
Z_{gluon}(q^2)\sim(q^2)^{2\kappa}\,,\,Z_{ghost}(q^2)\sim(q^2)^{-\kappa},
\end{equation}
with $\kappa\sim0.595$, has been extensively used in subsequent works (see \cite{fischer06} for a 
recent review). As shown in figure 1 of \cite{brasil}, these power laws 
are only valid for very low momenta, $q<200 MeV$ (see also \cite{fispaw}).
To test this solution of
the DSE with lattice QCD, using a symmetric lattice, 
it would require a lattice volume much 
larger than a typical present day simulation (see, for example,
 \cite{sternbeck}). 

Large asymmetric lattices, in the form $L_s^3 \times L_t$, with $L_t \gg L_s$, 
give us a possibility to test these power laws on the lattice. In this paper we briefly report on the results \cite{sardenha, dublin, finvol, nossoprd, madrid, brasil, tucson} obtained by us, considering 
large asymmetric lattices with $L_s=8,10,\ldots,18$ and $L_t=256$, about 
the infrared behaviour of the gluon and ghost propagators and the strong 
running coupling defined from these propagators. Despite the 
finite volume effects caused by the small spatial extension, the large 
temporal size of these lattices allow to access to momenta as low as 48 MeV.

In what concerns the gluon propagator, our results \cite{nossoprd} show that 
the propagator dependence on the spatial volume is smooth. Indeed, for the 
smallest momenta, the bare gluon propagator decreases with the lattice volume, 
and increases for higher momenta. The values of the 
infrared exponent extracted from our lattices increase with the lattice 
volume. Although almost all values of $\kappa$ are below 0.5 (see table 1 in 
\cite{nossoprd}), we obtain, by extrapolating the $\kappa$ values to the 
infinite volume,  $\kappa$ values above 0.5, with a weigthed mean of the 
various estimations giving $\bar{\kappa}_{\infty}= 0.5246(46)$.

\begin{figure}[!t]
\psfrag{EIXOX}{{\footnotesize $q(GeV)$}}
\psfrag{EIXOY}{{\footnotesize $D(q^2)$}}
   \begin{minipage}[b]{0.45\textwidth}
   \centering
   \includegraphics[width=6.5cm]{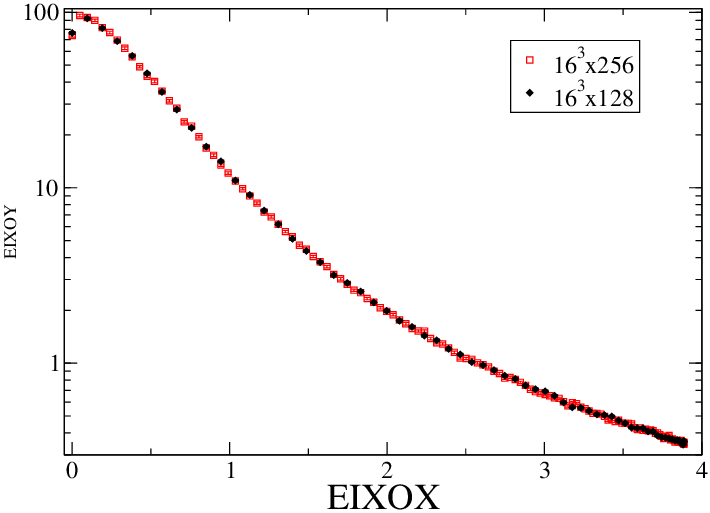}
  \end{minipage} 
\hspace{0.08\textwidth}
  \begin{minipage}[b]{0.45\textwidth}
  \centering
  \includegraphics[width=6.5cm]{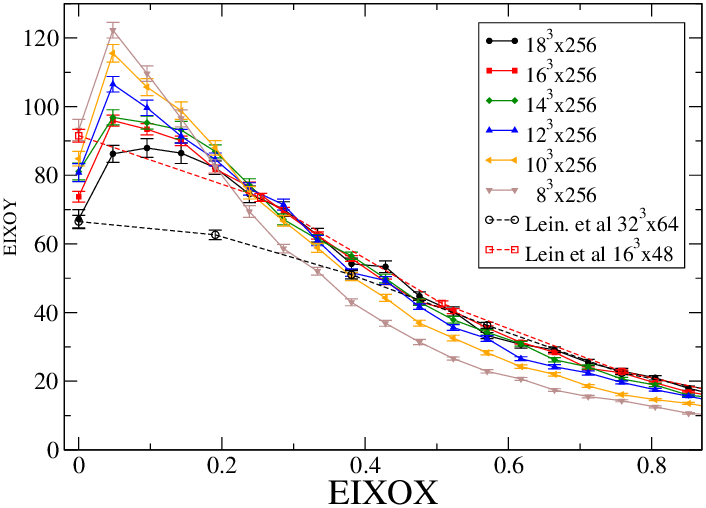}
  \end{minipage} 
\caption{On the left, the gluon propagator for $16^3\times 128$ 
and $16^3\times 256$ lattices, considering 
only pure temporal momenta. Note the logarithmic scale 
in the vertical axis. On the right, the gluon propagator for 
all lattices $L_{s}^{3}\times256$. For comparisation, we also show the 
$16^3\times 48$ and $32^3\times 64$ propagators computed in \cite{lein99}. } 
\label{asygluon}
\end{figure}

Considering the gluon propagator as a function of the spatial volume, we can 
also extrapolate it, and  fit the obtained propagator to a pure power law. 
Going this way, we get values for $\kappa \in [0.49,0.53]$. Note that the 
lattice data favours the values in the right hand side of
this interval.

The reader should be also aware that fits to our data considering higher momenta and other model functions give higher values for $\kappa$ \cite{brasil,poster}.

Similarly to other studies, it is possible to use our gluon data to verify the positivity violation for the gluon propagator \cite{tucson,poster}.

We have also computed the ghost propagator and the strong coupling constant $\alpha_S(q^2)$ defined from these propagators, for our smallest lattices \cite{madrid}. Our lattice data for these quantities also show sizeable dependence on the spatial volume of the lattices involved in our calculations. We also have found visible Gribov copy effects in the ghost propagator as well as in the strong coupling constant.

\begin{figure}[!h]
\psfrag{EIXOX}{{\footnotesize $q(GeV)$}}
\psfrag{EIXOY}{{\footnotesize \hspace{-0.5cm}$Z_{ghost}(q^2)$}}
   \begin{minipage}[b]{0.45\textwidth}
   \centering
   \includegraphics[width=6.5cm]{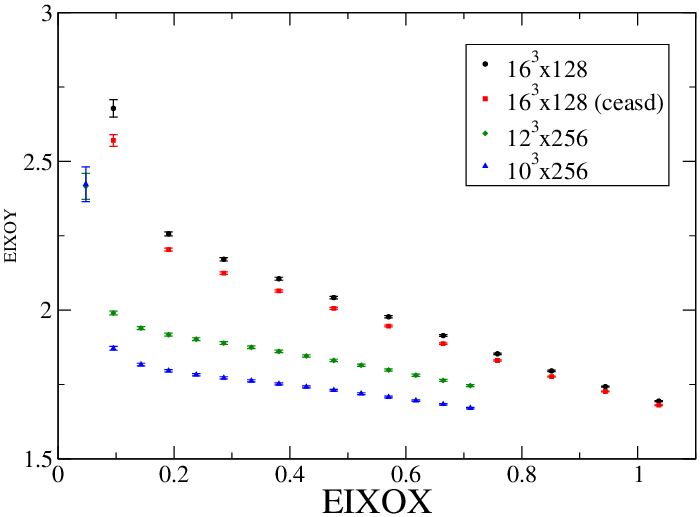}
  \end{minipage} 
\hspace{0.08\textwidth}
\psfrag{EIXOY}{{\footnotesize \hspace{-0.6cm}$\alpha(q^2)/\alpha(\mu^2)$}}
  \begin{minipage}[b]{0.45\textwidth}
  \centering
  \includegraphics[width=6.5cm]{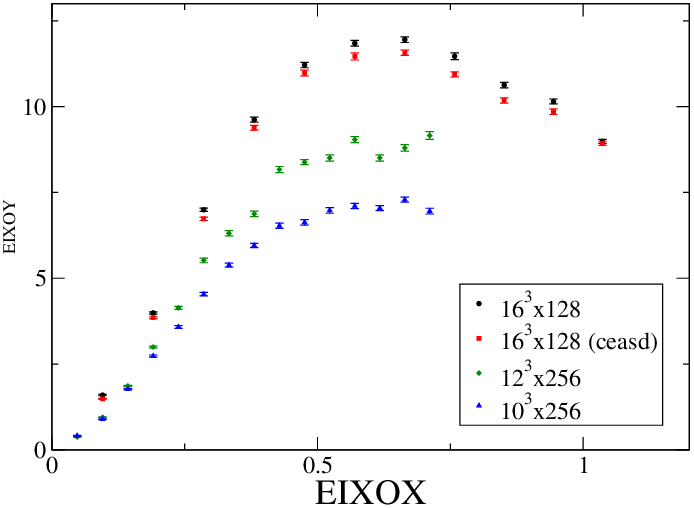}
  \end{minipage} 
\caption{ On the left, the bare ghost dressing function in the infrared region computed from a plane wave source. On the right, the  strong coupling constant. Here, we only consider pure temporal momenta.  \label{ghostalpha}} 
\end{figure}

Concerning the infrared behaviour of the ghost propagator, we were unable to extract an infrared exponent from our results. Possible reasons for this negative result can be either the finite volume effects associated to the small spatial volume of the lattices involved in the computation, or the lack of lattice data in the infrared region --- remember that the DSE ghost power law lacks validity well below 200 MeV. 

In the infrared region, $\alpha_S(q^2)$ shows a decreasing behaviour for the smallest momenta, in apparent contradiction with the continuum DSE prediction --- an infrared fixed point, but in agreement with other lattice studies \cite{sternbeck} and the solution of DSE on a torus \cite{dsetorus}. However, the reader should be aware that $\alpha_S(q^2)$, for the smallest momenta, seems to increase with the volume. 

In a near future, we will improve the statistics for our larger lattices and the extrapolations to the infinite volume limit. We also plan to perform simulations with larger lattices.


\vspace*{-0.2cm}
\begin{theacknowledgments}
This work was supported by FCT via grant 
SFRH/BD/10740/2002, and project POCI/FP/63436/2005.
\end{theacknowledgments}


\end{document}